\documentclass[preprint]{aastex701}

\begin{document}

\title{No Pulsar Detected in Reprocessed Archival Parkes Observations of SNR 1987A}

\author[0000-0002-2578-0360]{Fronefield Crawford}
\affiliation{Department of Physics and Astronomy, Franklin and Marshall College, P.O. Box 3003, Lancaster, PA 17604, USA}
\email[show]{fcrawfor@fandm.edu}  

\author{Haoyang Xu}
\affiliation{Emma Willard School, 285 Pawling Avenue, Troy, NY 12180, USA}
\email{rebeccahaoyangxu@gmail.com}

\begin{abstract}
We have reprocessed the available archival radio pulsar search observations of SNR 1987A taken with the Parkes 64-m telescope, some of which have not been previously published. We conducted a standard periodicity search on these data as well as a single pulse search at a range of dispersion measures. We found no convincing candidate signals, and we calculate flux density, luminosity, and single pulse fluence limits from these observations. The derived luminosity limits are comparable to the luminosities of three young, energetic pulsars (the Crab pulsar, PSR B0540$-$69, and PSR J0537$-$6910), and so we cannot rule out the existence of a pulsar in SNR 1987A with a similar radio luminosity. 
\end{abstract}


\section{Background and Motivation}

Searches for a radio pulsar in SNR 1987A in the Large Magellanic Cloud (LMC) commenced with the Parkes (``Murriyang'') 64-m radio telescope soon after the supernova was detected. \citet{m88} and \citet{m07} reported upper limits from several searches at a range of radio frequencies, going as high as 8 GHz. Later searches by  \citet{zdh+18} included a single pulse search in addition to a standard periodicity search. They reported the detection of several single pulse candidates, but none were convincingly associated with SNR 1987A. More recent search observations with MeerKAT by \citet{plg+24} were also unsuccessful. However, the reprocessing of previously searched data can sometimes reveal pulsars missed in prior attempts. A notable example is the discovery of more than a hundred new pulsars in the Parkes Multibeam Pulsar Survey by \citet{fsk+04} in reprocessing after the survey had been completed. In addition, some of the observations we have reprocessed here have apparently not been previously published.

\section{Data Analysis}

We downloaded all Parkes pulsar search observations targeting SNR 1987A that were available in the CSIRO Data Archive.\footnote{\url{https://data.csiro.au}} We found 6 separate observations between 2006 and 2019 which were part of Parkes project codes P505 and P834. Three of these observations had been split into subbands in the archive, and we processed each of these separately. Table \ref{tbl-1} shows the details of these observations. The observations from 2013 were previously analyzed and reported by \citet{zdh+18} (see their Table 1), but we were unable to find some other observations in their Table 1. Table 1 in \citet{m07} lists observations taken in late 2006, but they do not appear to include the 2006 observation we reprocessed. The 2019 observations in our table have not been previously reported.

We first searched for periodicities using PRESTO \citep{r01}. The data were flagged for radio interference then dedispersed at dispersion measures (DMs) from 0 to 1000 pc cm$^{-3}$. Each dedispersed time series was Fourier transformed, and each resulting power spectrum harmonically summed. Promising candidates with signal-to-noise (S/N) greater than 7 were refolded at periods and DMs near the candidate values. Our single pulse search used HEIMDALL \citep{bbb+12} to identify dispersed pulses at DMs from 0 to 1000 pc cm$^{-3}$. Boxcar matched filters were applied to each dedispersed time series, with widths ranging from 1 to 512 samples in powers of two. This corresponded to sensitivity to a maximum pulse width of $\sim 50$\,ms in each case. Pulses with S/N greater than 10 were then given to FETCH \citep{aab+20}, a deep‐learning based classifier that identifies astrophysical pulses. All pulses with a FETCH-assigned likelihood of being real greater than 50\% were inspected visually. 

\section{Results and Conclusions}

We found no convincing astrophysical signals above our thresholds in any of the observations. The single-pulse fluence limits $F_{\rm min}$ are given in Table \ref{tbl-1}, where a 1 ms pulse width is assumed. The corresponding values at 1400 MHz are also estimated, where a Crab-like spectral index for giant pulses of $-2.6$ is assumed \citep{mtb+17} . Table \ref{tbl-1} also lists flux density limits $S_{\rm min}$ from the periodicity searches assuming a 5\% pulsed duty cycle, and the scaled limits at 1400 MHz assumed a Crab pulsar spectral index of $-3.1$ for periodic emission \citep{mkk+00}. The (pseudo)luminosity limits $L_{\rm min}$ were determined by $L_{\rm min}  = S_{\rm min} d^{2}$, where a distance $d$ of 50 kpc to the LMC was used \citep{pgg+13}. 

We can compare our luminosity limits to the luminosities of three young, energetic pulsars that might share similar characteristics to a young pulsar in SNR 1987A: the Crab pulsar, PSR B0540$-$69, and PSR J0537$-$6910. Their 1400-MHz luminosities are 56, 32, and 25 mJy kpc$^{2}$, respectively, as listed in the ATNF Pulsar Catalog \citep{mht+05}.\footnote{\url{https://www.atnf.csiro.au/research/pulsar/psrcat/}. Note that the luminosity of PSR J0537$-$6910 is only an upper limit since pulsed radio emission has not been detected.} The 1400-MHz luminosity limits in Table \ref{tbl-1} span more than an order of magnitude, but the best of these limits are comparable to these pulsar luminosities. We therefore cannot significantly constrain or rule out the existence of a pulsar in SNR 1987A with a similar radio luminosity.

Future search observations might be more successful. First, next-generation instruments (e.g., the Square Kilometer Array) will have better raw sensitivity and therefore better limits. Second, SNR 1987A is expected to become increasingly transparent to radio emission at lower frequencies over time as the supernova expands. \citet{m88} indicated that the nebula ought to become optically thin at $\sim$~GHz frequencies after about a decade, with scattering no longer being a significant factor after this time. However, \citet{wsw+19} suggested that as of 2018, the plasma cutoff frequency was approximately 33 GHz. If this is the case, then this opacity could be preventing the detection of the pulsar at lower frequencies.

\tabletypesize{\tiny}
\begin{deluxetable}{lcccccccccccc}
\tablecaption{Reprocessed Parkes Search Observations of SNR 1987A from the CSIRO Archive}
\tablewidth{0pt}
\tablehead{
\colhead{Date} \vspace{-0.6cm} & \colhead{Rcvr} &  \colhead{Freq.} & \colhead{BW}    & \colhead{$N_{\rm chan}$} & \colhead{$t_{\rm samp}$} & \colhead{$T_{\rm int}$}  & \colhead{$S_{\rm min}$} & \colhead{$S_{\rm min}^{1400}$} & \colhead{$L_{\rm min}$}    & \colhead{$L_{\rm min}^{1400}$} & \colhead{SP $F_{\rm min}$} & \colhead{SP $F_{\rm min}^{1400}$} \\ \\
\colhead{}                     & \colhead{}     &  \colhead{(MHz)} & \colhead{(MHz)} & \colhead{}               &\colhead{($\mu$s)}            & \colhead{(hr)}           & \colhead{($\mu$Jy)}     & \colhead{$\mu$Jy)}  & \colhead{(mJy\,kpc$^{2}$)} & \colhead{(mJy\,kpc$^{2}$)}     & \colhead{(Jy\,ms)}         & \colhead{(Jy\,ms)} 
}
\startdata
2006/02/19 & MULTI       & 1390  &  256 &  512 & 125 & 1.18 &  64 & 63 & 160 &  157 & 0.807 &  0.792 \\
\hline
2013/05/08 & 1050CM      &  732  &   64 &  512 &  96 & 1.50 & 227 & 30 & 567 &   76 & 0.608 &  0.113 \\
           &             & 3100  & 1024 &  512 & 100 & 1.50 &  39 & 455 &  97 & 1138 & 4.428 & 34.982 \\
\\
\hline
2013/05/24 & MULTI       & 1369  &  256 &  512 & 100 & 6.50 & 20 & 19 &  50 &   46 & 0.564 &  0.532 \\
\hline
2013/05/25 & MULTI       & 1369  &  256 &  512 &  96 & 7.11 & 19 & 18 &  47 &   44 & 0.564 &  0.532 \\
\hline
2019/01/28       & UWL         & 1369  &  256 &  512 &  64 & 4.25 & 22 & 21 &  56 &   52 & 0.514 &  0.485 \\
           &             & 1024  &  640 & 4096 & 128 & 6.35 &  12 & 4 &  29 &   11 & 0.153 &  0.068 \\
           &             & 1856  & 1024 & 4096 &  64 & 1.67 &  18 & 43 &  45 &  107 & 0.567 &  1.181 \\
\hline
2019/03/31 & UWL         & 1856  & 1024 & 4096 &  64 & 5.83 &  10 & 23 &  24 &   57 & 0.567 &  1.181 \\
           &             & 3200  & 1664 & 3328 &  64 & 7.92 &   6 & 83 &  16 &  209 & 1.835 & 15.742 \\
\enddata 
\tabletypesize{}
\tablecomments{Multiple entries on the same date represent a single observation with subbands that were processed separately. In some cases, only some of the raw files from the archive were usable, yielding different total integration times for observations on the same date. Receivers: MULTI = Multibeam; 1050CM = 10/50 cm dual-band; UWL = Ultra-Wideband Low.} 
\label{tbl-1}
\end{deluxetable}

\bibliography{cx25}{}
\bibliographystyle{aasjournalv7}

\end{document}